\documentclass{webofc}
\usepackage[varg]{txfonts}   

\usepackage{color,ulem,cancel}
\usepackage{tabularx}
\usepackage{float}

\begin{document}

\title{How does dark matter affect compact star properties and high density constraints of strongly interacting matter\\
}

\author{
        \firstname{Violetta} \lastname{Sagun}\inst{1}\fnsep\thanks{\email{violetta.sagun@uc.pt}} \and
\firstname{Edoardo} \lastname{Giangrandi}\inst{1}\fnsep\thanks{\email{eg@student.uc.pt}} 
\and
        \firstname{Oleksii} \lastname{Ivanytskyi}\inst{2}\fnsep\thanks{\email{oleksii.ivanytskyi@uwr.edu.pl}}  \and
         \firstname{Constança} \lastname{Providência}\inst{1}\fnsep\thanks{\email{cp@uc.pt}}
              \and
                     \firstname{Tim} \lastname{Dietrich}\inst{3,4}\fnsep\thanks{\email{tim.dietrich@aei.mpg.de}}
}

\institute{CFisUC, Department of Physics, University of Coimbra, Rua Larga P-3004-516, Coimbra, Portugal
\and
        Institute of Theoretical Physics, University of Wroclaw, 50-204
Wroclaw, Poland
\and
Institut für Physik und Astronomie, Universität Potsdam, Haus 28,
Karl-Liebknecht-Str. 24/25, Potsdam, Germany
\and
Max Planck Institute for Gravitational Physics (Albert Einstein Institute), Am Mühlenberg 1, Potsdam, Germany
          }

\abstract{%
We study the impact of asymmetric bosonic dark matter on neutron star properties, including possible changes of tidal deformability, maximum mass, radius, and matter distribution inside the star. The conditions at which dark matter particles tend to condensate in the star's core or create an extended halo are presented. We show that dark matter condensed in a core leads to a decrease of the total gravitational mass and tidal deformability compared to a pure baryonic star, which we will perceive as an effective softening of the equation of state. On the other hand, the presence of a dark matter halo increases those observable quantities. Thus, observational data on compact stars could be affected by accumulated dark matter and, consequently, constraints we put on strongly interacting matter at high densities. To confirm the presence of dark matter in the compact star's interior, and to break the degeneracy between the effect of accumulated dark matter and strongly interacting matter properties at high densities, several astrophysical and GW tests are proposed.

}
\maketitle
\setlength{\footskip}{3.60004pt}
\section{Introduction}
\label{intro}
After almost nine decades of searches for the nature of dark matter (DM), its exact properties still remains unveiled. Part of these searches is based on recoil experiments that try to measure the direct evidence of the interaction between DM and baryonic matter (BM) and obtain new constraints on the DM properties, e.g. mass and cross section with nuclei \citep{Klasen:2015uma}. Cosmological probes, e.g. the cosmic microwave background (CMB), Bullet Cluster, galaxies and large-scale structure formation, gravitational lensing provide constraints on DM from macroscopic observables \citep{Clowe_2006,Boddy:2022knd}. In addition, astrophysical observations come along to further test up-to-date DM models on objects, such as stars. Compact stars, in particular, may contain a sizeable amount of accumulated DM, that has been accumulated throughout all evolution steps, i.e. DM-admixed proto-cloud, main-sequence phase, supernova explosion, and, finally, equilibrated  neutron stars (NS). As a consequence, the NS properties, such as mass and radius, may be affected by the presence of DM.

As it was already shown in Refs.~\cite{2019JCAP...07..012N,PhysRevD.102.063028,Ellis:2017jgp}, depending on the DM mass, its relative fraction inside a star and self-interaction strength, the NS properties can be severely modified. Thus, light DM particles, e.g. with mass $m_\chi\sim100$ MeV, tend to create a dilute halo around a NS, while heavier DM particles, e.g. with mass $m_\chi\sim1$ GeV, are confined in the inner NS region, forming a DM core. Hence, the outermost radius of DM-admixed NSs depends on the DM properties: in a halo configuration, the DM radius will exceed the baryonic one; whereas in a core scenario, the outermost radius will coincide with the visible one of BM.

As a consequence of the DM presence, the observable gravitational mass and radius of NSs will be modified. As a matter of fact, as shown in Refs.~\cite{2019JCAP...07..012N,2021PhRvD.104f3028D,Sagun:2021oml,Rafiei_Karkevandi_2022}, heavy DM particles lead to a decrease of the mass and radius, making a star more compact. The same effect of increasing the compactness can be caused by the appearance of new degrees of freedom, e.g.\ $\Lambda$-hyperons, or a phase transition to quark-gluon plasma in the core of a compact star \cite{Ivanytskyi:2022mlk}. In the other scenario, when DM is distributed in an extended halo, higher masses can be reached.

Here we show an existing degeneracy between the effect of accumulated DM and strongly interacting matter properties at high densities. The presence of DM could mimic a softening/stiffening of the baryonic equation of state (EoS), and, therefore, the constraints we put on strongly interacting matter at high densities.

Based on the observations of the four heaviest pulsars PSR J0348+0432 with a mass of 2.01$\pm$0.04 M$_\odot$ \cite{PSRj03480432Article}, PSR J0740+6620 with 2.14$^{+0.10}_{-0.09}$ M$_\odot$ \cite{PSRJ0740+6620Article}, PSR J1810+1744 of $2.13\pm 0.04$ M$_{\odot}$ \citep{Romani:2021xmb}, and PSR J0952-0607 of $2.35\pm 0.17$ M$_{\odot}$ \citep{Romani:2022jhd} one can impose a lower limit on the mass of DM-admixed NSs. These constraint can then be further extended through the combination with data from the first binary NS merger detection GW170817 \cite{LIGOScientific:2017vwq}, which provided important constraints on the tidal deformability $\Lambda(1.4\mathrm{M}_\odot)\leq800$ at 90\% confidence level.

\section{Equation of state of dark and baryon matter}
\label{sec-EoS}
In this work, we model DM as a complex scalar field $\chi$ with mass $m_\chi$, carrying a conserved charge. Such bosonic DM, at a sufficiently low temperature, forms a Bose-Einstein Condensate (BEC) and, when no interactions are considered, it exerts zero pressure. Thus, it could lead to gravitational instabilities \citep{Kouvaris}. In order to stabilize DM-admixed NSs, we consider a repulsive self-interaction, mediated by the real vector $\omega^\mu$ field with mass $m_\omega$. Such a system can be described by the following minimal Lagrangian
\begin{equation}
\label{eq.FullLagrangian}
\mathcal{L} = (D_\mu\chi)^*D^\mu\chi-m_\chi^2 \chi^*\chi -\frac{\Omega_{\mu\nu}\Omega^{\mu\nu}}{4}+\frac{m_\omega^2\omega_\mu\omega^\mu}{2},
\end{equation}
where the scalar field is coupled to the $\omega^\mu$ field through the covariant derivative $D^\mu=\partial^\mu-ig\omega^\mu$. Here, $g$ is the corresponding Yukawa coupling constant and $\Omega_{\mu\nu}=\partial_\mu\omega_\nu-\partial_\nu\omega_\mu$.

\begin{figure}[h]
\centering
\includegraphics[width=13cm,clip]{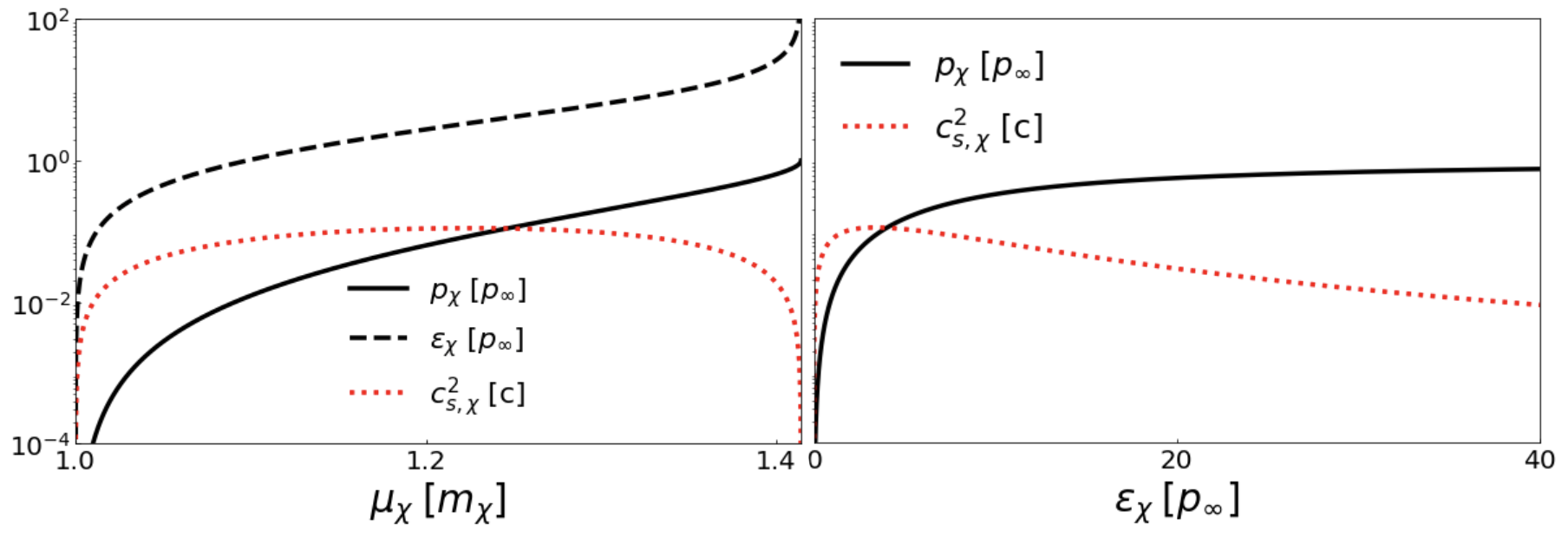}
\caption{\textbf{Left panel:} Scaled pressure $p_\chi/p_\infty$ (black solid curve), energy density $\varepsilon_\chi/p_\infty$ (black dashed curve) and speed of sound squared $c_{s,\chi}^2$ (red dotted curve) of DM as functions of its chemical potential $\mu_\chi$ given in units of $m_\chi$. {\bf Right panel:}
Scaled pressure $p_\chi/p_\infty$ (black solid curve) and speed of sound $c_{s,\chi}^2$ (red dotted curve) of DM as functions of scaled energy density $\varepsilon_\chi$ given in units of $p_\infty$.}
\label{fig-EoS}       
\end{figure}

In this work, we assume vanishing DM temperature, where thermal fluctuations are completely suppressed. Thus, it is possible to apply the mean field approximation in order to obtain the corresponding expressions for pressure and energy density, which  in the Grand Canonical Ensemble (GCE) can be written as 
\begin{equation}\label{eq:eos}
p_\chi=\frac{m_I^2}{4}
\left(m_\chi^2-\mu_\chi\sqrt{2m_\chi^2-\mu_\chi^2}\right),\ \ \
\varepsilon_\chi=\frac{m_I^2}{4}
\left(\frac{\mu_\chi^3}{\sqrt{2m_\chi^2-\mu_\chi^2}}-m_\chi^2\right).
\end{equation}
Here $\mu_\chi$ is the DM chemical potential, and $m_I\equiv m_\omega / g$ is the interaction strength. As it can be seen from Eq.~\eqref{eq:eos}, the DM chemical potential is limited to the interval [1,$\sqrt{2}$]$m_\chi$. At $\mu_\chi\rightarrow\sqrt{2}m_\chi$ energy density diverges, while pressure saturates to $p_\infty\equiv m_I^2m_\chi^2/4$. This leads to vanishing compressibility of DM at asymptotically high densities. The same behaviour holds for the speed of sound $c_s^2\equiv \partial p_\chi / \partial \varepsilon_\chi$, as it can be seen from Fig.\ref{fig-EoS}. As a matter of fact, such behaviour corresponds to gravitationally unstable matter at high densities for any considered value of the interaction strength or mass of the DM particles.

To study the effect of DM regardless of the BM properties, we consider three different hadronic EoSs with different stiffness. One of them is the IST EoS, being a parametric EoS of the NS matter accounting for a hard-core repulsion between nucleons. This EoS reproduces the nuclear matter properties, includes the nuclear liquid-gas phase transition and its critical point \cite{Sagun:2016nlv}, proton flow constraints \cite{Ivanytskyi:2017pkt}, and hadron yields created in heavy-ion collisions \cite{Sagun:2017eye}. In this work, we consider the parameter set B proposed in Ref. \cite{Sagun:2020qvc}, whereas the crust is described by the polytropic EoS with adiabatic index $\gamma=4/3$. Furthermore, we also consider the DD2 EoS with and without $\Lambda$ hyperons \cite{Typel1999, Typel2009}. This EoS is the mean-field relativistic nuclear model with density-dependent couplings. Hereby, the density dependence of the couplings of $\Lambda$ hyperons to $\sigma$, $\omega$ and $\rho$ mesons is assumed to be the same as the one of nucleons. All parameters are fitted to reproduce the nuclear matter ground state properties. The BPS EoS \cite{bps} is used for the outer crust, the inner crust is modeled within a Thomas-Fermi framework, taking the DD2 EoS as the underlying model and considering pasta phases with different geometries as discussed in Refs. \cite{grill14,Fortin2016}.

Observations of the Bullet Cluster \cite{Clowe_2006} provide useful constraints on the interaction between DM and BM, showing that the cross-section is many orders of magnitude lower than the typical nuclear one. Hence, with a good approximation, we assume no interaction between DM and BM, apart from the gravitational one. As a consequence, the stress-energy tensors of the two fluids are conserved separately. As it has been shown by Ivanytskyi et al. \cite{PhysRevD.102.063028} a formulation of the DM EoS in the GCE gives an advantage of having two chemical potentials of the BM and DM components that scale proportionally. This significantly simplifies solving two coupled Tolman-Oppenheimer-Volkoff (TOV) equations, which are split due to independent conservation of the corresponding energy-momentum tensors. Thus, for (i=D,B)
\begin{equation}\label{TOV}
\frac{dp_i}{dr}=-\frac{(\epsilon_i +p_i)(M_\mathrm{tot}+4\pi r^3p_\mathrm{tot})}{r^2\left(1-{2M_\mathrm{tot}}/{r}\right)}.
\end{equation}
Here $M_{tot}=M_D+M_B$ and $p_{tot}=p_D+p_B$ are the total gravitational mass and pressure, respectively. Once integrated the TOV equations, it is possible to quantify the amount of DM in the configuration through the individual masses via the DM fraction $f_\chi = \frac{M_D}{M_{tot}}$.

\section{Results}

\begin{figure}[h]\centering
\setkeys{Gin}{width=1.2\linewidth}
\begin{tabularx}{\linewidth}{XXX}
\includegraphics{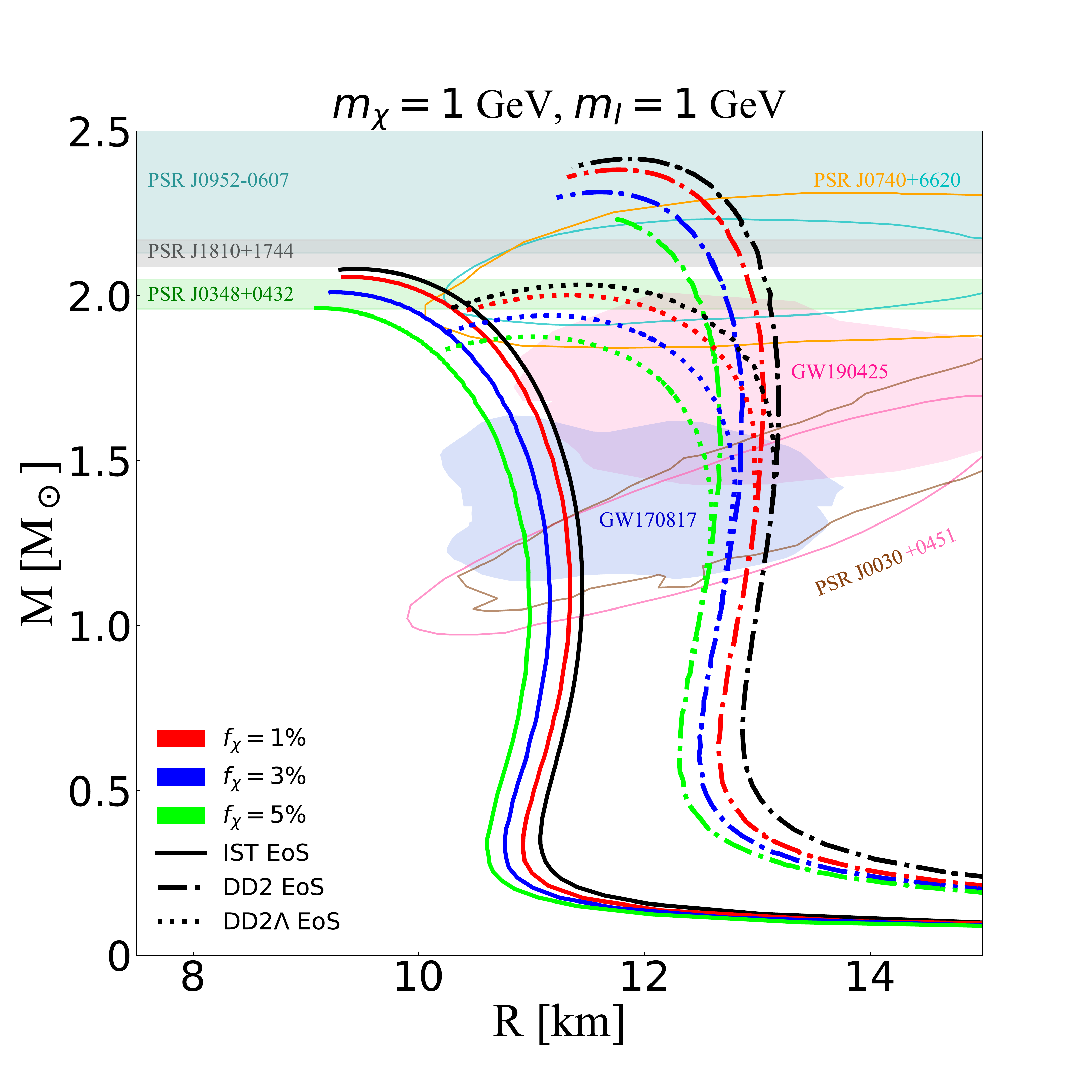}
    &
\includegraphics{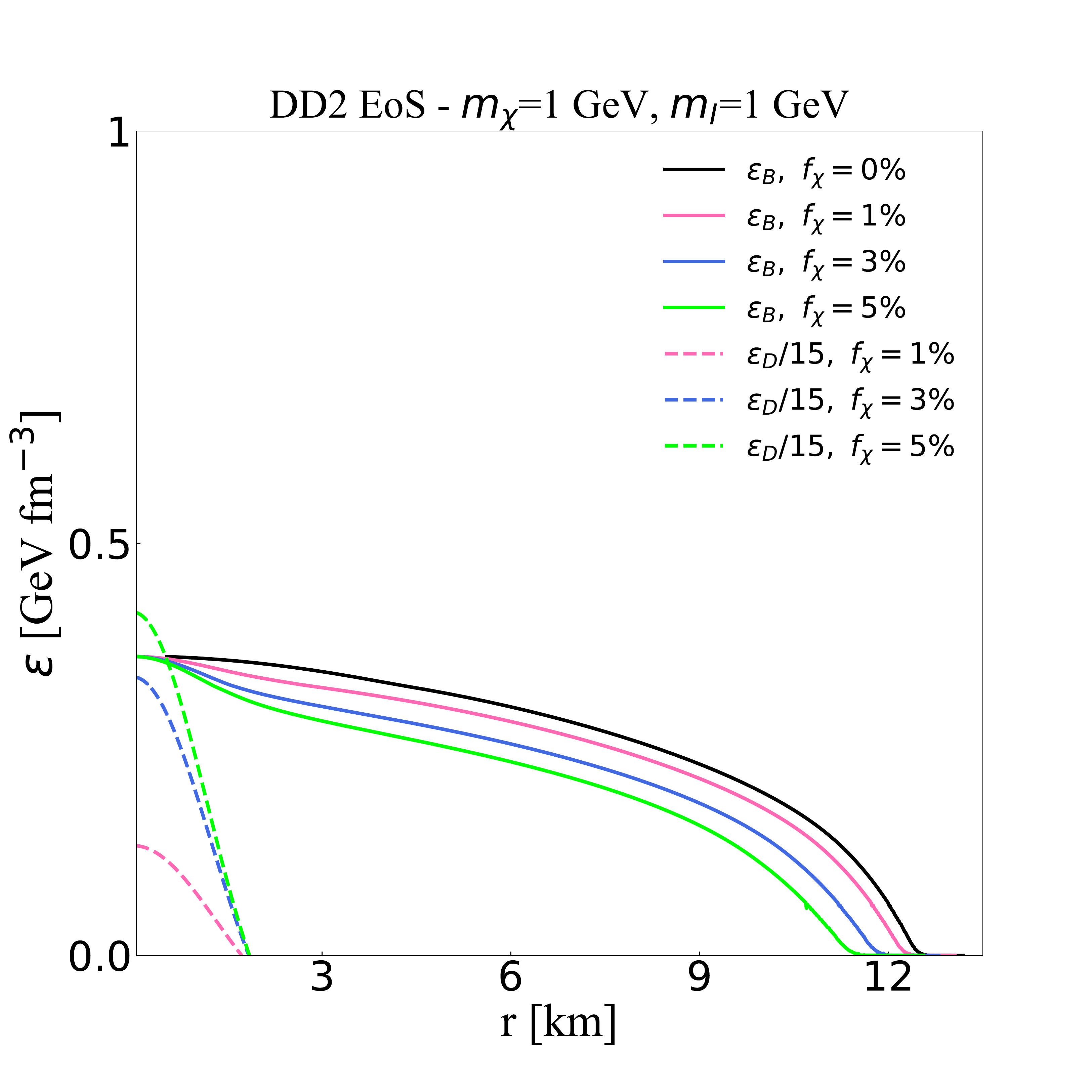}
    &
\includegraphics{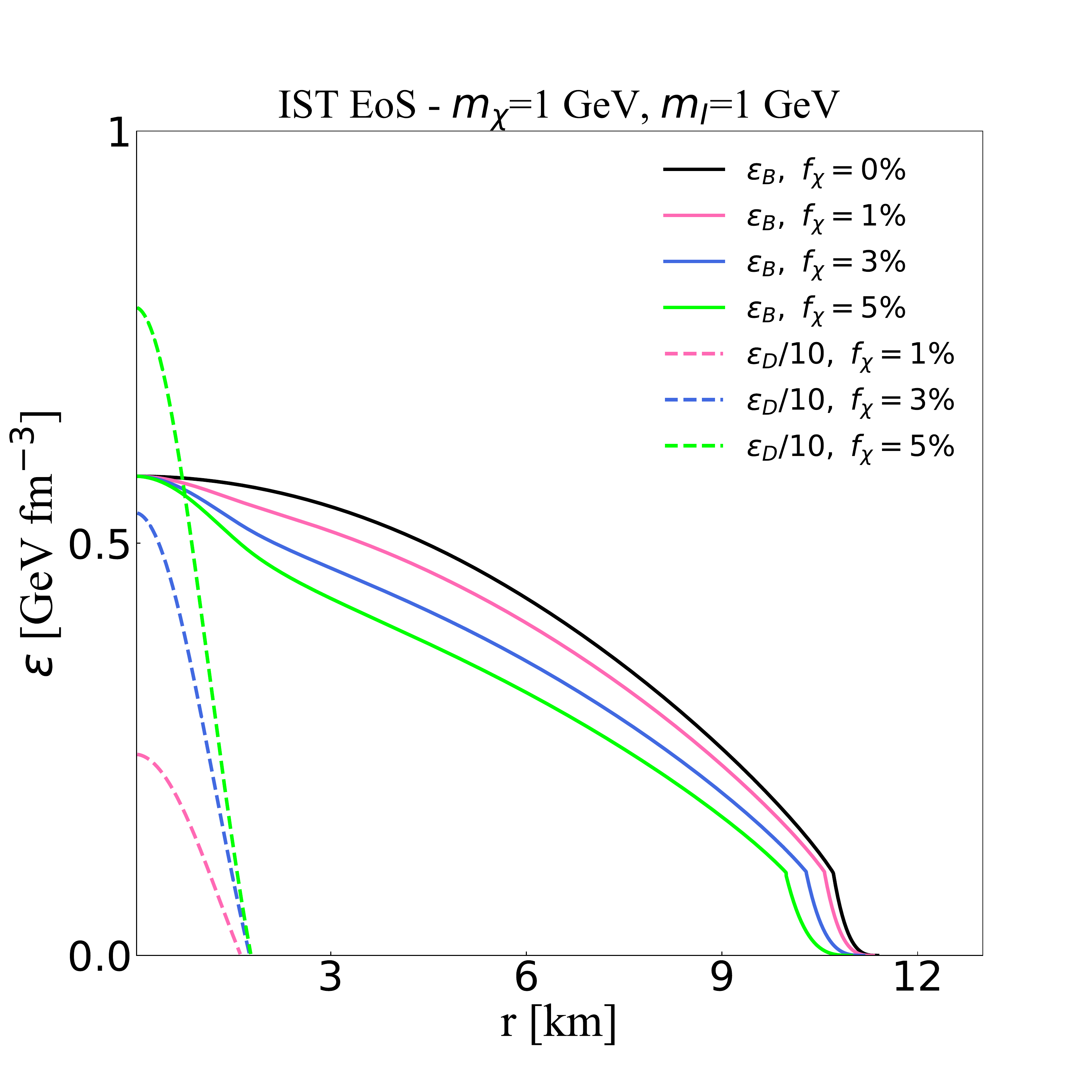}
    \end{tabularx}
        \begin{tabularx}{\linewidth}{XXX}
\includegraphics{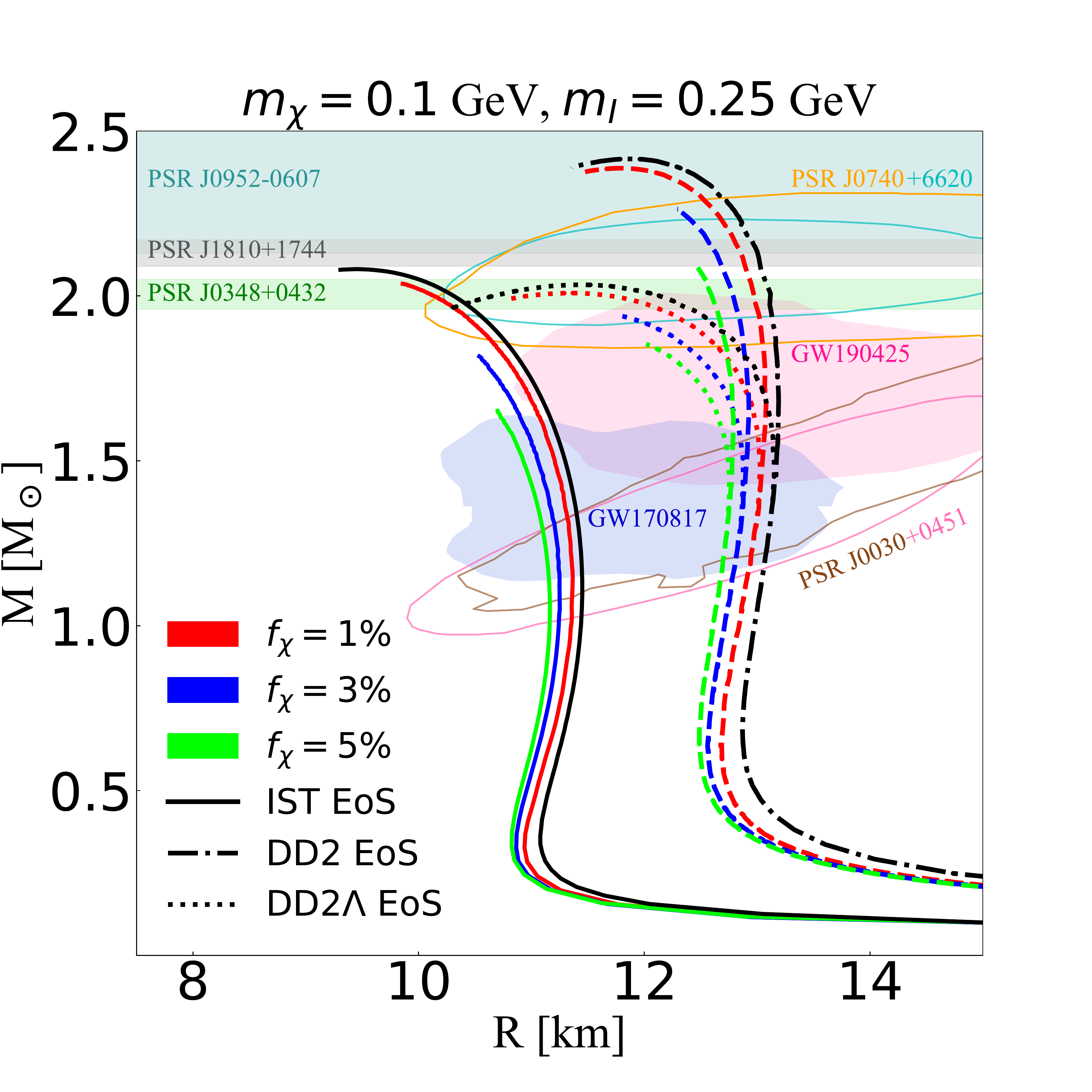}
    &
\includegraphics{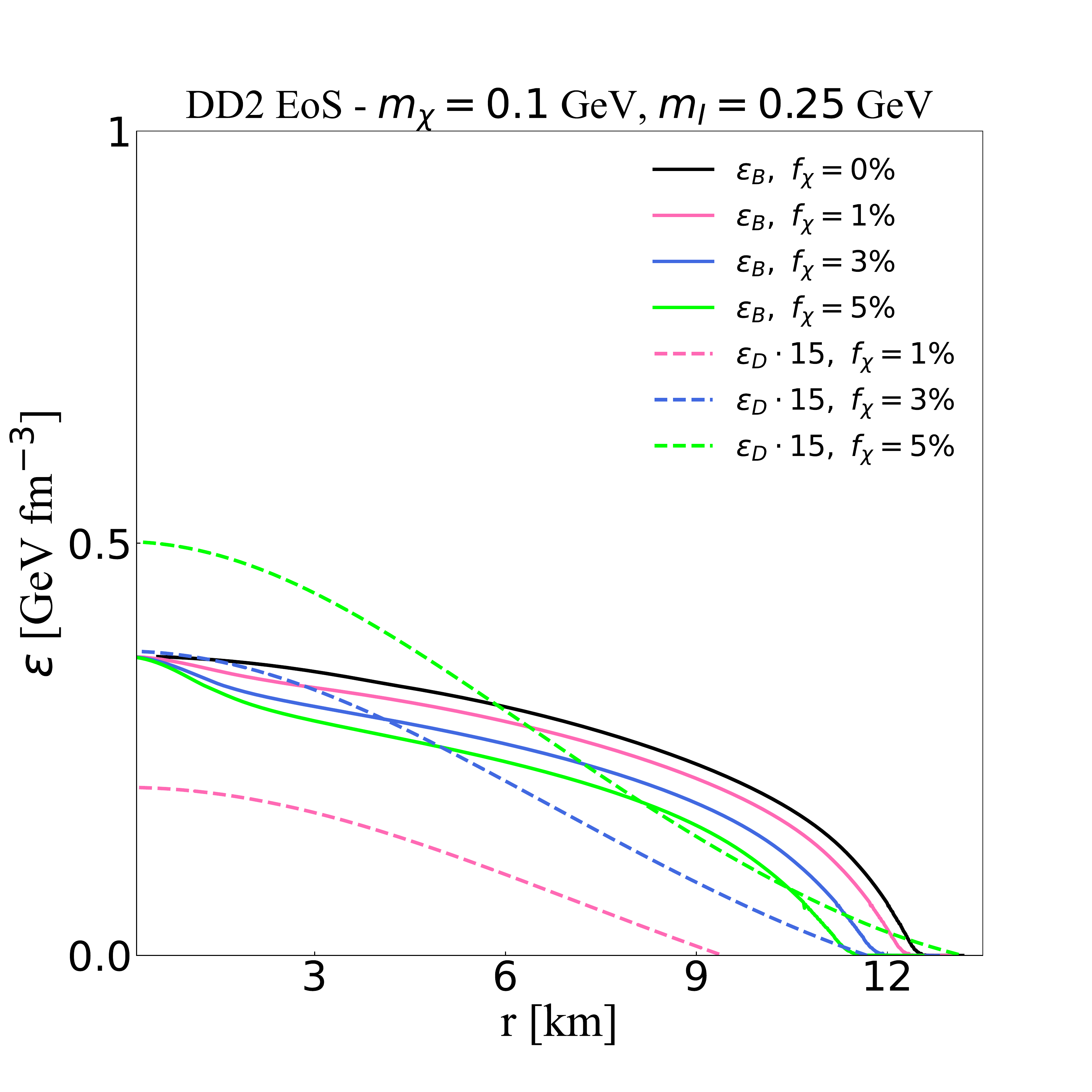}
    &
\includegraphics{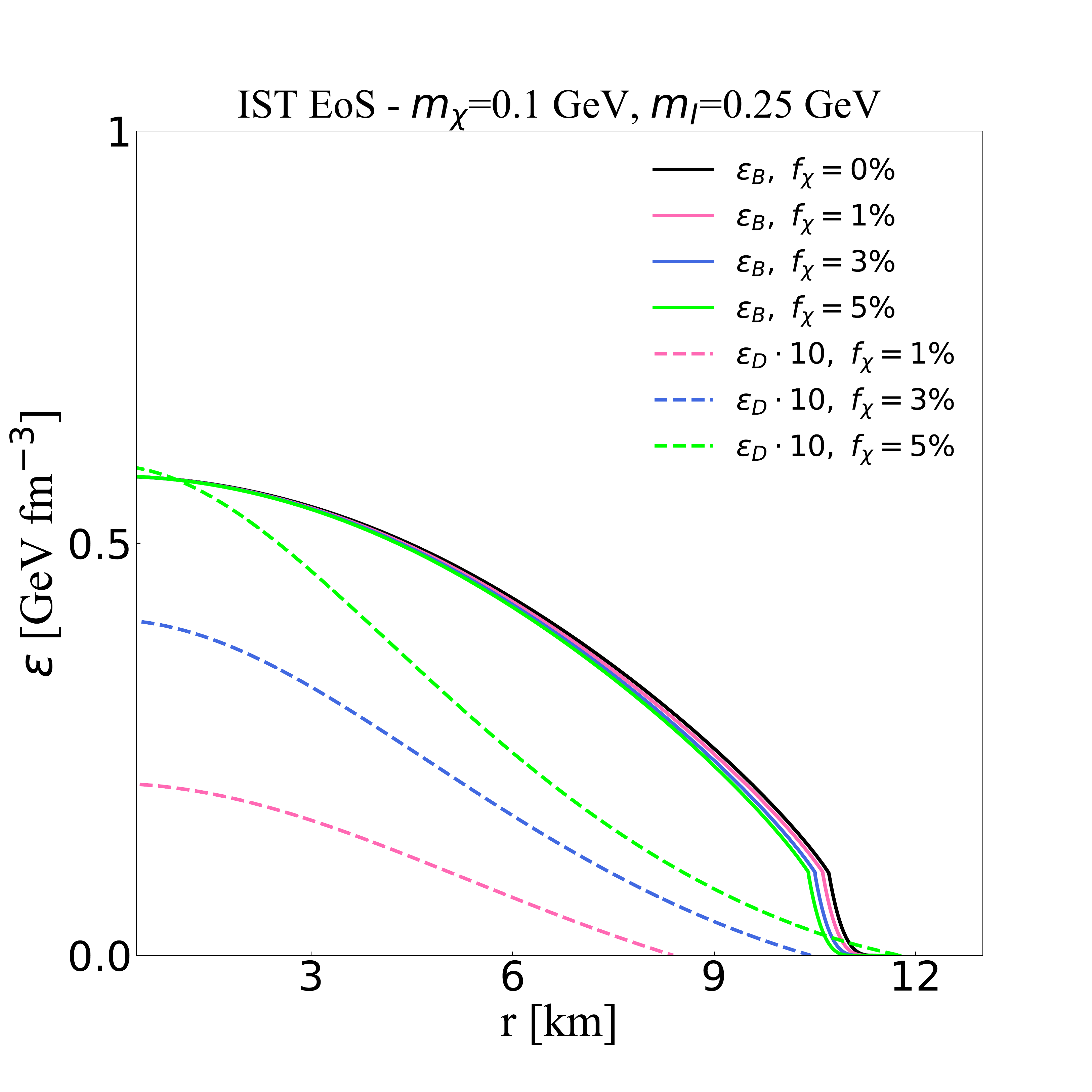}
\end{tabularx}
        \begin{tabularx}{\linewidth}{XXX}
\includegraphics{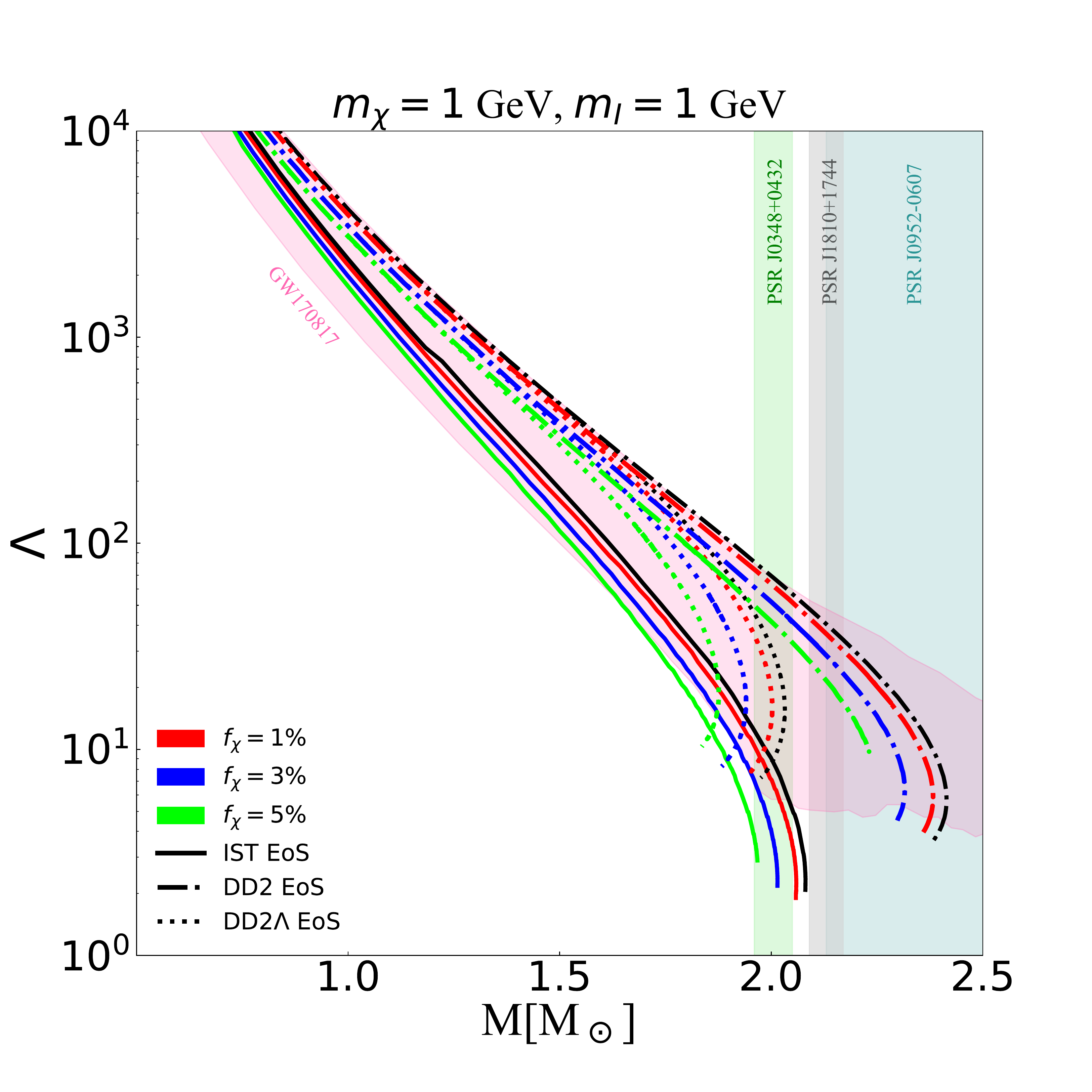}
    &
\includegraphics{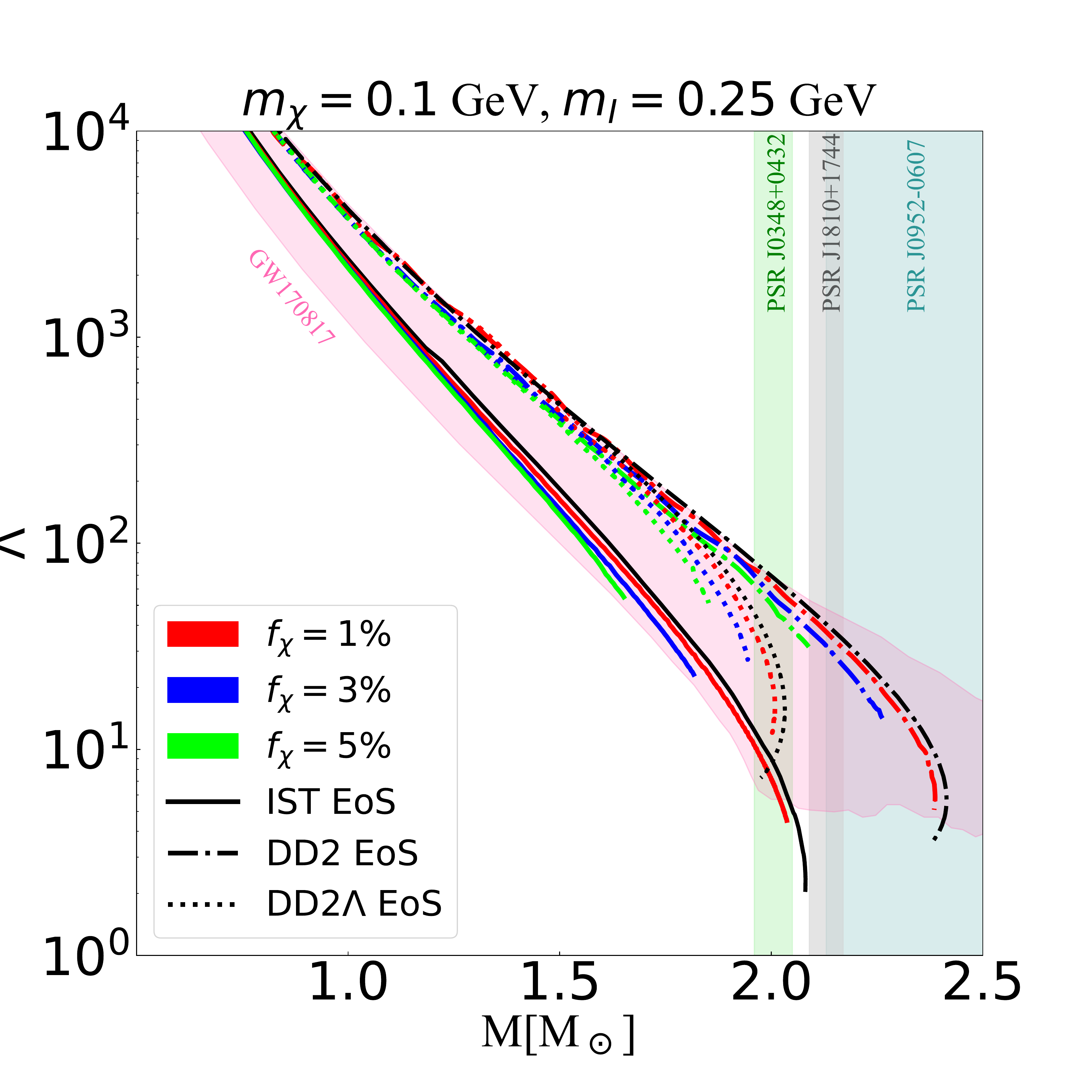}
    &
\includegraphics{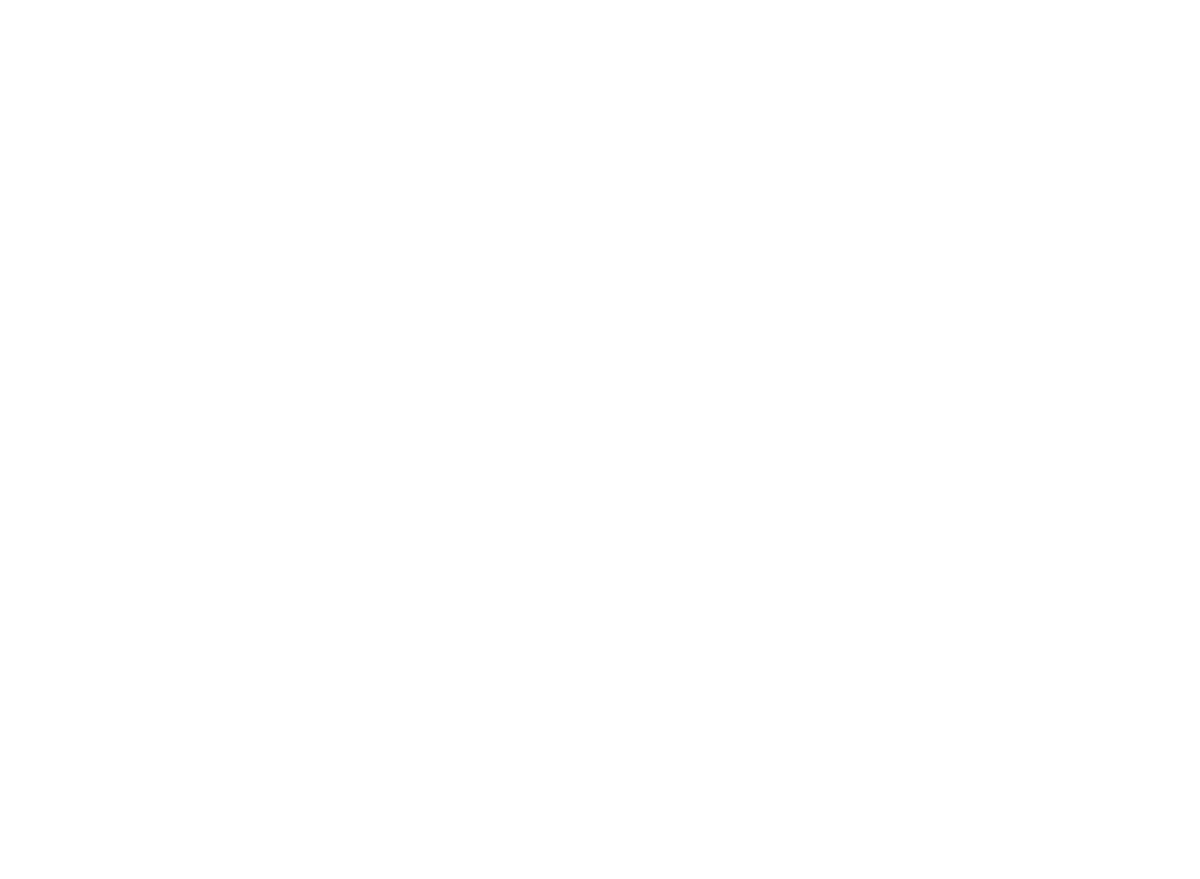}
\end{tabularx}
\caption{ {\bf Upper row:}
Total gravitational mass of the DM-admixed NSs as a function of its visible radius $R$ (left panel). Black curves correspond to pure BM stars described by the IST EoS, DD2 EoS and DD2 EoS with hyperons (dubbed as DD2$\Lambda$). Red, blue, and green colours depict relative DM fractions equal to 1\%, 3\%, and 5\%, correspondingly.
Green, gray, and teal bands represent 1$\sigma$ constraints on the mass of, respectively, PSR J0348+0432 \citep{PSRj03480432Article}, PSR J1810+1744 \citep{Romani:2021xmb}, and PSR J0952-0607 \citep{Romani:2022jhd}. Pink and beige contours show the NICER measurements of PSR J0030+0451 \citep{Riley:2019yda,Miller_2019}, while orange and blue contours depict the PSR J0740+6620 measurements \citep{Miller:2021qha,Riley:2021pdl}. LIGO-Virgo observations of GW170817 \citep{Abbott_2018} and GW190425 \citep{LIGOScientific:2020aai} binary NS mergers are shown in blue and magenta.
Energy density profiles for BM (solid curves) and DM (dotted curves) components are shown for the DD2 EoS (middle panel) and IST EoS (right panel). While the solid black curve depicts the profile for a 1.4 M$_\odot$ NS, the other profiles were obtained considering the same central baryonic chemical potential as of the pure baryonic star and adding DM to the configurations. Both panels were obtained for $m_{\chi}$=1 GeV, $m_{I}$=1 GeV.
{\bf Middle row:} The same as on the upper row, but for $m_{\chi}$=0.1 GeV, $m_{I}$=0.25 GeV.
{\bf Lower row:}
Tidal deformability as a function of total gravitational mass calculated for pure BM stars (black curves) and DM-admixed NSs for $m_{\chi}$=1 GeV, $m_{I}$=1 GeV (left panel) and $m_{\chi}$=0.1 GeV, $m_{I}$=0.25 GeV (middle panel) with relative DM fractions 1\%, 3\%, and 5\%, in red, blue, and green, correspondingly. The magenta area visualizes the constraints obtained from GW170817 \citep{Abbott_2018}. Other curves and bands follow the notation of the top left panel. 
}
\label{fig:MRprofiles}
\end{figure}

The mass-radius curves obtained by the integration of the two-fluid TOV equations are shown in Fig.~\ref{fig:MRprofiles}. The black curves represent pure baryonic NSs, where no DM is present. As it can be seen, both DD2 EoSs coincide up to 1.4 M$_\odot$ where an onset of the $\Lambda$ hyperons occurs. Appearance of $\Lambda$ hyperons softens the EoS, leading to lower maximum mass. However, regardless of the considered BM EoS, massive DM particles, e.g. $m_\chi=1$ GeV, form a dense core in the inner regions of a star, creating a further gravitational pull, and leading to more compact NS configurations, as it can be seen from the upper row of Fig.~\ref{fig:MRprofiles}. Such DM cores can be clearly seen from the energy density profiles, shown in the upper row. The DM component vanishes at $R_{D}\sim2.5$ km, being the DM core radius. Inside the sphere of this radius, the value of the DM energy density, $\varepsilon_D$, is an order of magnitude higher than the baryonic one, representing the main component in the NS interior.

\begin{figure}[h]\centering
\setkeys{Gin}{width=1\linewidth}
\begin{tabularx}{\linewidth}{XXX}
\includegraphics{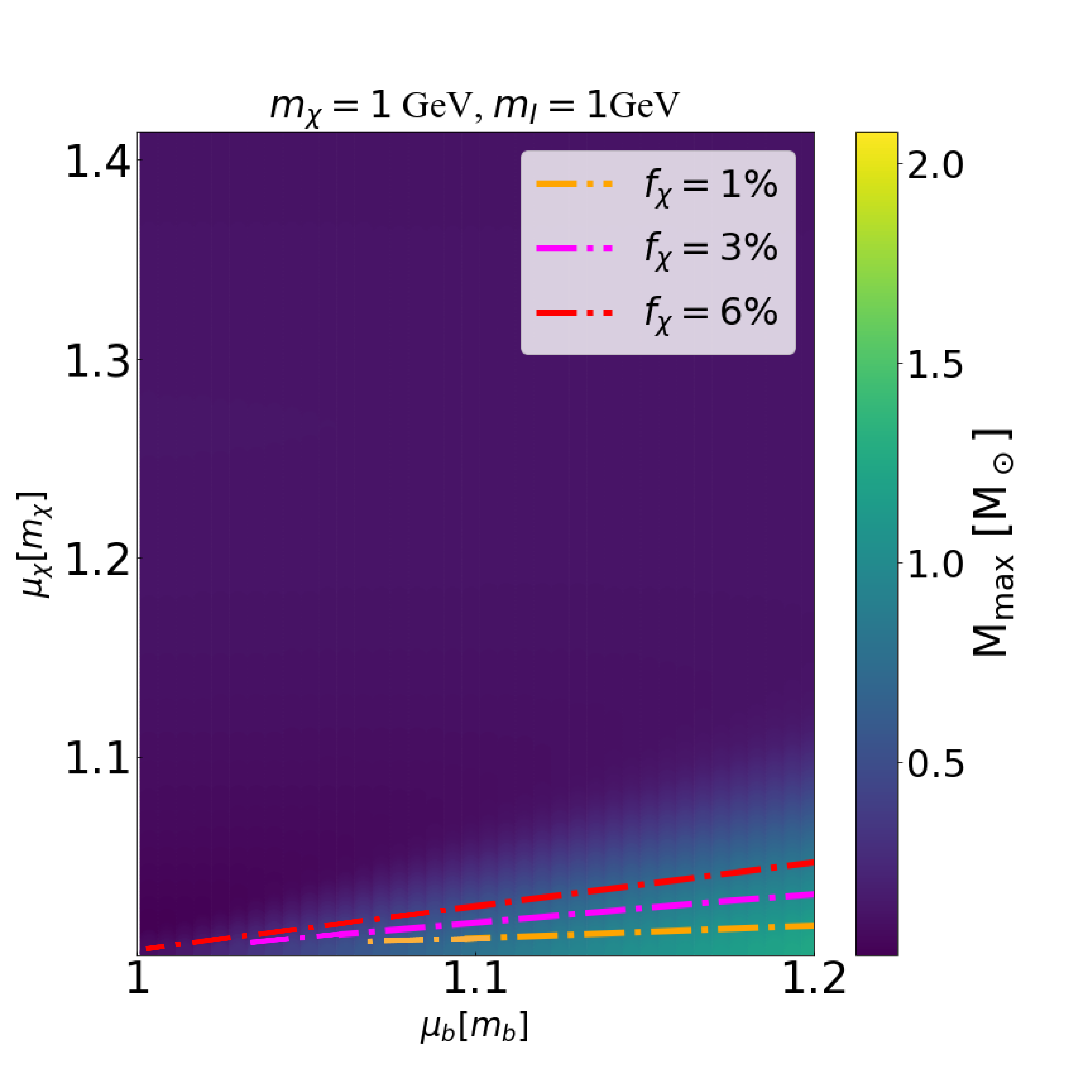}
    &
\includegraphics{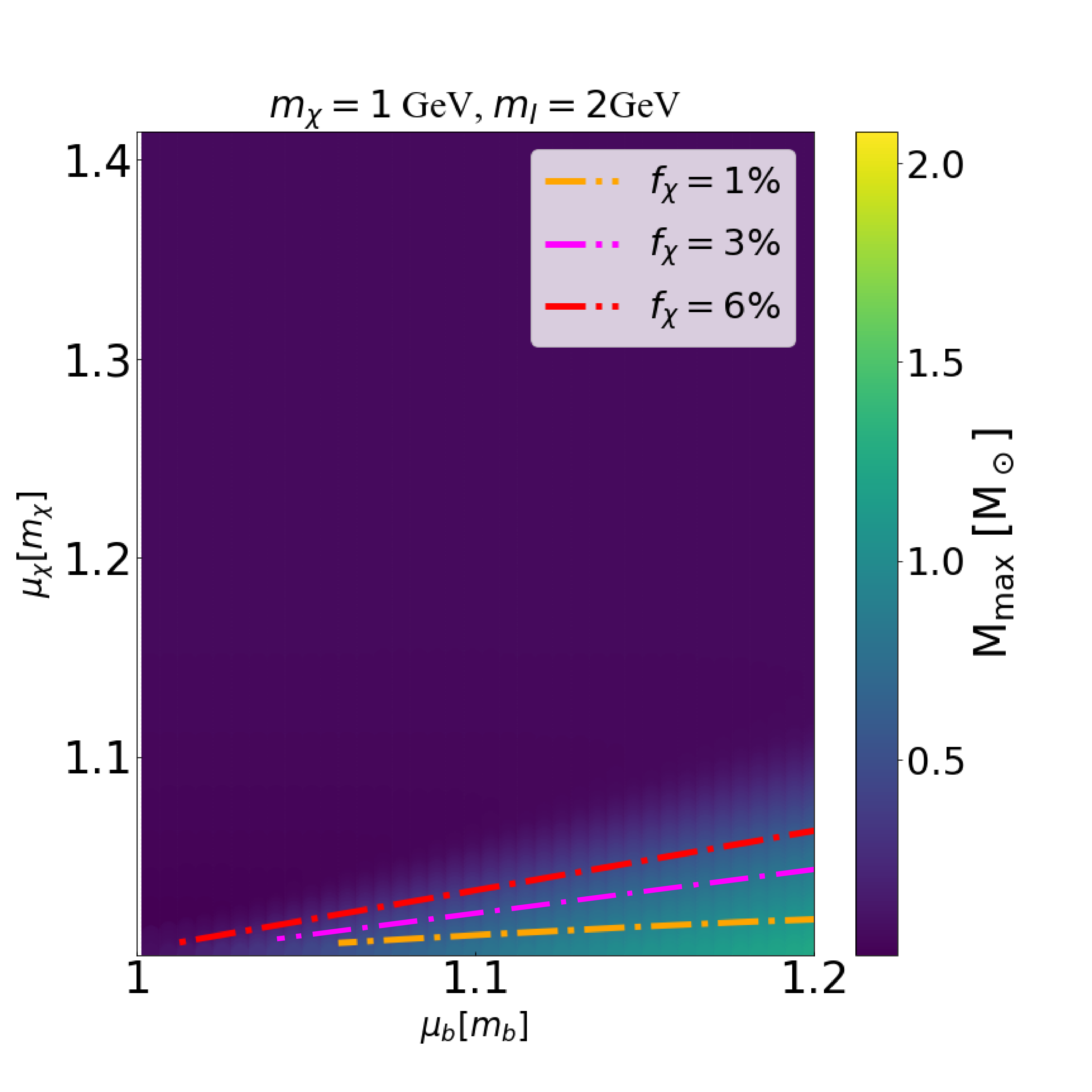}
    \end{tabularx}
        \begin{tabularx}{\linewidth}{XXX}
\includegraphics{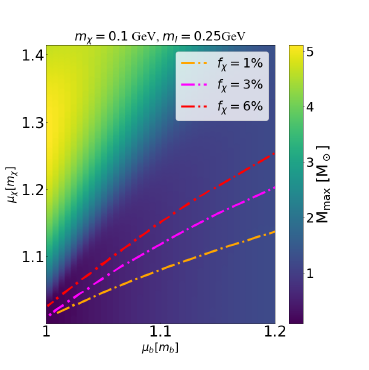}
    &
\includegraphics{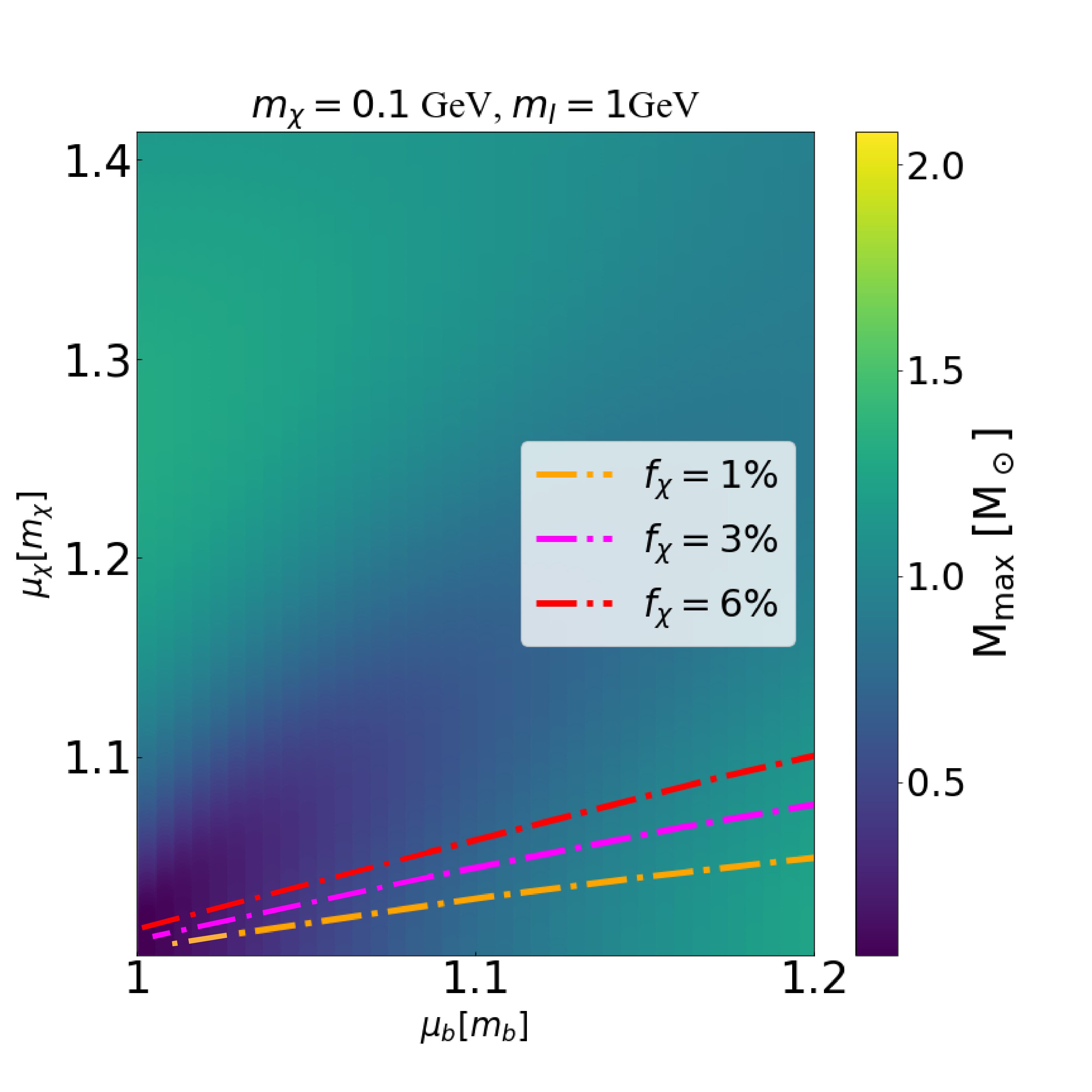}
\end{tabularx}
\caption{Parameter space in the $\mu_\chi -\mu_b$ plane calculated for the IST EoS and different values of model parameters: $m_{\chi}$=1 GeV, $m_{I}$=q GeV (upper left panel), $m_{\chi}$=1 GeV, $m_{I}$=2 GeV (upper right panel), $m_{\chi}$=0.1 GeV, $m_{I}$=0.25 GeV (lower left panel), $m_{\chi}$=0.1 GeV, $m_{I}$=1 GeV (lower right panel). The color represents the total maximum gravitational mass of DM-admixed NSs. The yellow, magenta, and red dash-dotted curves correspond to the DM fraction, 1\%, 3\%, and 6\%.}
\label{fig:scan}
\end{figure}

On the other hand, lighter DM particles, e.g. $m_\chi\sim$ 0.1 GeV, tend to form diluted halos embedding the NSs. Indeed, looking at the profiles in the middle row of Fig.~\ref{fig:MRprofiles}, the DM energy density is non-zero at a radius of the BM component, forming a DM halo with $R_D>R_B$. However, for light DM particles, the chemical potential constraint plays a huge role. When $\mu_\chi\rightarrow\sqrt{2}m_\chi$, DM pressure asymptotically approaches the constant value $p_\infty$, not being able to exert sufficient pressure to avoid mechanical instabilities. As it can be seen on the middle left panel of Fig.~\ref{fig:MRprofiles}, the $M$-$R$ curves suddenly end due to a black hole (BH) formation, leading to an absence of stable DM-admixed NS configurations.

Fig.~\ref{fig:scan} shows an interplay between the DM and BM chemical potentials, and their effect on the maximum gravitational mass of DM-admixed NSs. The scan is performed for the IST EoS and four different sets of the DM particle mass and interaction scale. The yellow, magenta, and red dash-dotted curves represent fixed DM fractions of 1\%, 3\%, and 6\%. The lower left panel of Fig.~\ref{fig:scan} exhibits much higher values of the total maximum gravitational mass of DM-admixed NSs that correspond to halo configurations. For low values of baryonic chemical potential and high dark chemical potentials, the model allows the existence of exotic dark objects with the DM fraction higher than 95\%. Due to this, the total maximum gravitational mass of these systems reaches up to 5M$_\odot$.

The modifications in the profiles lead to significant changes in the quadrupole deformabilities $\Lambda=2/3 k (R_\mathrm{outermost}/M_\mathrm{tot})^5$, where $R_\mathrm{outermost}=\mathrm{max}(R_D,R_B)$ is the outermost radius of the DM-admixed NS and $k$ is the Love number \cite{Hinderer_2008}, which encodes the main EoS information. On the lower row of  Fig.~\ref{fig:MRprofiles} we show the $\Lambda-M_\mathrm{tot}$ curves using the two-fluid approach presented in details in Ref.  \cite{Giangrandi:2022wht}, and evaluated by varying the DM fraction. As a result, the greater the amount of DM, the lower the NS tidal deformabilities, regardless of the baryonic EoS. This can be seen as one of the effects of DM, further compressing the stars they get harder to be deformed and tidally disrupted.

\section{Conclusions}\label{sec:Conclusions}
Here we analyze the effect of self-interacting bosonic DM on the NS properties, such as mass, radius, profile and tidal deformability. Depending on DM particle's properties, i.e. particle mass and interaction strength, two different configurations can be expected: DM core or DM halo. An accumulated DM in the inner NS region leads to the lower maximum gravitational mass and tidal deformability, compared to the pure baryonic stars. These star configurations are more compact and harder to be deformed. On the other hand, when DM is distributed in a diluted halo, the outermost radius is defined by DM, leading to larger tidal deformabilities. Thus, considering the LIGO/Virgo Collaboration upper bound on the tidal deformability parameter $\Lambda_{1.4}<800$ we can conclude that GW170817
disfavours extended DM halos embedding the NSs in a binary system in the host NGC 4993 galaxy.

As it was mentioned, accumulated DM could affect observables in a way that it will resemble stiffening (in the case of a halo configuration) or softening (in the case of a core configuration) of the BM EoS at high density. This kind of degeneracy could lead to misleading results regarding the strongly interacting matter properties at high density. As was thoroughly discussed in Ref. \citep{Giangrandi:2022wht} the presence of DM could leave several smoking gun evidences. Among them, it is worth mentioning measurements of mass, radius, and moment of inertia of NSs at a different distance from the Galactic center with few-\%-accuracy. Thus, we expect a higher DM fraction inside compact stars towards the Galactic center, and, therefore, a stronger DM impact. A correlation between the maximum gravitational mass, radius, or any other observable quantities of compact stars with the distance from the Galactic center will put a strong constraint on DM properties and be an independent probe free from the BM EoS uncertainties. DM could leave an imprint on a gravitational wave signal from NS-NS and NS-BH mergers that will exhibit as the presence of supplementary peak(s) in the gravitational wave spectrum, exotic waveform or modification of the kilonova ejecta. Moreover, another effects such as gravitational-lensing or alteration of the pulsar pulse profile due to the extra light-bending in a dark halo could be present.

\section*{Acknowledgments}
The work is supported by national funds from FCT – Fundação para a Ciência e a Tecnologia, I.P., within the Project No. EXPL/FIS-AST/0735/2021. E.G., C.P. and V.S. acknowledge the support from FCT within the Projects No. UIDB/04564/2020, UIDP/04564/2020. E.G. also acknowledges the support from the Project No. PRT/BD/152267/2021. The work of O.I. was supported by the Polish National Science Center under the grant No. 2019/33/BST/03059.
\bibliography{ConfXVproceeding}{}

\begin{thebibliography}{33}

\bibitem{Klasen:2015uma}
M.~Klasen, M.~Pohl, G.~Sigl, Prog. Part. Nucl. Phys. \textbf{85}, 1 (2015),
  \texttt{1507.03800}

\bibitem{Clowe_2006}
D.~Clowe, M.~Brada{\v{c}}, A.H. Gonzalez, M.~Markevitch, S.W. Randall,
  C.~Jones, D.~Zaritsky, The Astrophysical Journal \textbf{648}, L109 (2006)

\bibitem{Boddy:2022knd}
K.K. Boddy et~al., JHEAp \textbf{35}, 112 (2022), \texttt{2203.06380}

\bibitem{2019JCAP...07..012N}
A.E. Nelson, S.~Reddy, D.~Zhou, J. Cosm. Astrop. Phys. \textbf{2019}, 012
  (2019), \texttt{1803.03266}

\bibitem{PhysRevD.102.063028}
O.~Ivanytskyi, V.~Sagun, I.~Lopes, Phys. Rev. D \textbf{102}, 063028 (2020)

\bibitem{Ellis:2017jgp}
J.~Ellis, A.~Hektor, G.~H{\"u}tsi, K.~Kannike, L.~Marzola, M.~Raidal,
  V.~Vaskonen, Phys. Lett. B \textbf{781}, 607 (2018), \texttt{1710.05540}

\bibitem{2021PhRvD.104f3028D}
H.C. Das, A.~Kumar, S.K. Patra, Phys. Rev. D \textbf{104}, 063028 (2021),
  \texttt{2109.01853}

\bibitem{Sagun:2021oml}
V.~Sagun, E.~Giangrandi, O.~Ivanytskyi, I.~Lopes, K.~Bugaev, PoS
  \textbf{PANIC2021}, 313 (2022), \texttt{2111.13289}

\bibitem{Rafiei_Karkevandi_2022}
D.R. Karkevandi, S.~Shakeri, V.~Sagun, O.~Ivanytskyi, Physical Review D
  \textbf{105} (2022)

\bibitem{Ivanytskyi:2022mlk}
O.~Ivanytskyi, D.~Blaschke, T.~Fischer, A.~Bauswein, \emph{{Early quark
  deconfinement in compact star astrophysics and heavy-ion collisions}}, in
  \emph{{29th International Conference on Ultra-relativistic Nucleus-Nucleus
  Collisions}} (2022), \texttt{2208.09085}

\bibitem{PSRj03480432Article}
J.~{Antoniadis}, P.C.C. {Freire}, N.~{Wex}, T.M. {Tauris}, R.S. {Lynch}, M.H.
  {van Kerkwijk}, M.~{Kramer}, C.~{Bassa}, V.S. {Dhillon}, T.~{Driebe} et~al.,
  Science \textbf{340}, 448 (2013), \texttt{1304.6875}

\bibitem{PSRJ0740+6620Article}
H.T. Cromartie, E.~Fonseca, S.M. Ransom, P.B. Demorest, Z.~Arzoumanian,
  H.~Blumer, P.R. Brook, M.E. DeCesar, T.~Dolch, J.A. Ellis et~al., Nature
  Astronomy \textbf{4}, 72–76 (2019)

\bibitem{Romani:2021xmb}
R.W. Romani, D.~Kandel, A.V. Filippenko, T.G. Brink, W.~Zheng, Astrophys. J.
  Lett. \textbf{908}, L46 (2021), \texttt{2101.09822}

\bibitem{Romani:2022jhd}
R.W. Romani, D.~Kandel, A.V. Filippenko, T.G. Brink, W.~Zheng, Astrophys. J.
  Lett. \textbf{934}, L18 (2022), \texttt{2207.05124}

\bibitem{LIGOScientific:2017vwq}
B.P. Abbott et~al. (LIGO Scientific, Virgo), Phys. Rev. Lett. \textbf{119},
  161101 (2017), \texttt{1710.05832}

\bibitem{Kouvaris}
C.~Kouvaris, \emph{The dark side of neutron stars} (2013),
  \urlstyle{tt}\url{https://arxiv.org/abs/1308.3222}

\bibitem{Sagun:2016nlv}
V.V. Sagun, K.A. Bugaiev, A.I. Ivanytskyi, D.R. Oliinychenko, I.N. Mishustin,
  EPJ Web Conf. \textbf{137}, 09007 (2017), \texttt{1611.07071}

\bibitem{Ivanytskyi:2017pkt}
A.I. Ivanytskyi, K.A. Bugaev, V.V. Sagun, L.V. Bravina, E.E. Zabrodin, Phys.
  Rev. C \textbf{97}, 064905 (2018), \texttt{1710.08218}

\bibitem{Sagun:2017eye}
V.V. Sagun, K.A. Bugaev, A.I. Ivanytskyi, I.P. Yakimenko, E.G. Nikonov, A.V.
  Taranenko, C.~Greiner, D.B. Blaschke, G.M. Zinovjev, Eur. Phys. J. A
  \textbf{54}, 100 (2018), \texttt{1703.00049}

\bibitem{Sagun:2020qvc}
V.~Sagun, G.~Panotopoulos, I.~Lopes, Phys. Rev. D \textbf{101}, 063025 (2020),
  \texttt{2002.12209}

\bibitem{Typel1999}
S.~Typel, H.H. Wolter, Nucl. Phys. A \textbf{656}, 331 (1999)

\bibitem{Typel2009}
S.~Typel, G.~Ropke, T.~Klahn, D.~Blaschke, H.H. Wolter, Phys. Rev. C
  \textbf{81}, 015803 (2010), \texttt{0908.2344}

\bibitem{bps}
G.~{Baym}, C.~{Pethick}, P.~{Sutherland}, Astrophys. J. \textbf{170}, 299
  (1971)

\bibitem{grill14}
F.~Grill, H.~Pais, C.~Provid\^encia, I.~Vida\~na, S.S. Avancini, Phys. Rev. C
  \textbf{90}, 045803 (2014)

\bibitem{Fortin2016}
M.~Fortin, C.~Providencia, A.R. Raduta, F.~Gulminelli, J.L. Zdunik, P.~Haensel,
  M.~Bejger, Phys. Rev. C \textbf{94}, 035804 (2016), \texttt{1604.01944}

\bibitem{Riley:2019yda}
T.E. Riley et~al., Astrophys. J. Lett. \textbf{887}, L21 (2019),
  \texttt{1912.05702}

\bibitem{Miller_2019}
M.C. Miller, F.K. Lamb, A.J. Dittmann, S.~Bogdanov, Z.~Arzoumanian, K.C.
  Gendreau, S.~Guillot, A.K. Harding, W.C.G. Ho, J.M. Lattimer et~al., The
  Astrophysical Journal \textbf{887}, L24 (2019)

\bibitem{Miller:2021qha}
M.C. Miller et~al., Astrophys. J. Lett. \textbf{918}, L28 (2021),
  \texttt{2105.06979}

\bibitem{Riley:2021pdl}
T.E. Riley et~al., Astrophys. J. Lett. \textbf{918}, L27 (2021),
  \texttt{2105.06980}

\bibitem{Abbott_2018}
B.P. Abbott et~al. (The LIGO Scientific Collaboration and the Virgo
  Collaboration), Phys. Rev. Lett. \textbf{121}, 1611012013Sci...340..448A
  (2018)

\bibitem{LIGOScientific:2020aai}
B.P. Abbott et~al. (LIGO Scientific, Virgo), Astrophys. J. Lett. \textbf{892},
  L3 (2020), \texttt{2001.01761}

\bibitem{Hinderer_2008}
T.~Hinderer, The Astrophysical Journal \textbf{677}, 1216 (2008)

\bibitem{Giangrandi:2022wht}
E.~Giangrandi, V.~Sagun, O.~Ivanytskyi, C.~Provid\^encia, T.~Dietrich (2022),
  \texttt{2209.10905}

\end{thebibliography}

\end{document}